\documentclass[final,3p,times,twocolumn]{elsarticle}

\usepackage{amssymb}
\usepackage{amsmath}
\usepackage{amsthm}
\usepackage{color}
\biboptions{numbers,sort&compress}
\usepackage{lineno}

\begin{document}

\begin{frontmatter}

\title{Calibration of liquid argon detector with $^{83m}Kr$ and $^{22}Na$ in different drift field}

\author[ihep,ucas] {Wei-Xing Xiong \corref{cor1}}
\author[ihep,lab]{Meng-Yun Guan}
\author[ihep] {Chang-Gen Yang}
\author[ihep,ucas]{Peng Zhang}
\author[ihep,lab]{Jin-Chang Liu}
\author[ihep,lab]{Cong Guo}
\author[ihep,lab]{Yu-Ting Wei}
\author[ihep,ucas,lab]{You-Yu Gan}
\author[ihep,lab]{Qing Zhao}
\author[ihep,ncwc] {Jia-Jun Li}

\address[ihep]{Institute of High Energy Physics, Beijing 100049, China}
\address[ucas] {University of Chinese Academy of Sciences, Beijing 100049, China}
\address[ncwc] {North China Electric Power University, Beijing 100096, China}
\begin{abstract}
  $^{83m}Kr$ and $^{22}Na$ have been used in calibrating a liquid argon (LAr) detector.$^{83m}Kr$ atoms are produced through the decay of $^{83}Rb$ and introduced into the LAr detector through the circulating purification system. The light yield reaches 7.26$\pm$0.02 photonelectrons/keV for 41.5keV from $^{83m}Kr$ and 7.66$\pm$0.01 photonelectrons/keV for the 511keV from $^{22}Na$, as a comparison. The light yield varies with the drift electric field from 50 to 200V/cm have been also reported. After stopping fill, the decay rate of $^{83m}Kr$ with a fitted half-life of 1.83$\pm$0.11 h, which is consistent with the reported value of 1.83$\pm$0.02 h.

\end{abstract}


\end{frontmatter}


\section{Introduction}
\label{Introduction}
Dark matter has been considered to be a particle in one or more forms, which makes up 23\% of the mass-energy density of the universe~\cite{Planck_2013_results}. Weakly Interacting Massive Particle(WIMP) is a well-motivated galactic dark matter candidate, which is the subject of many direct dark matter search experiments~\cite{Xenon100,PandaxII,Darkside,WArP,DEAP,CDMS}. Liquefied noble gases are widely used as targets in low background search experiments, particularly in direct dark matter search experiments where a WIMP may scatter elastically from a nucleus to produce a nuclear recoil, a dark matter particle of mass in the range of 10 to 1000 GeV$c^{-2}$ would give typical recoil energies in the range of 1 to 100 keV~\cite{RecoilE}. Several WIMP-nucleon cross-section limits have been set in recent years using liquid argon and xenon detectors~\cite{xenon1t,LUX,ThePandaX,TheDarkSide}, and several larger, more sensitive argon and xenon detectors are currently under construction~\cite{DS20K,PandaX4T,DARWIN}. As liquid noble-gas detectors get larger, self-shielding will render it increasingly difficult to illuminate the central volume of the liquid with external gamma rays, particularly at low energies. Therefore, a low energy radioactive source that can be distributed throughout the detector volume is of the essence for calibrating liquid noble-gas detectors to precisely determine its energy threshold and ultimate WIMP sensitivity.

\begin{figure}[!htb]
\begin{centering}
\includegraphics[width=.45\textwidth]{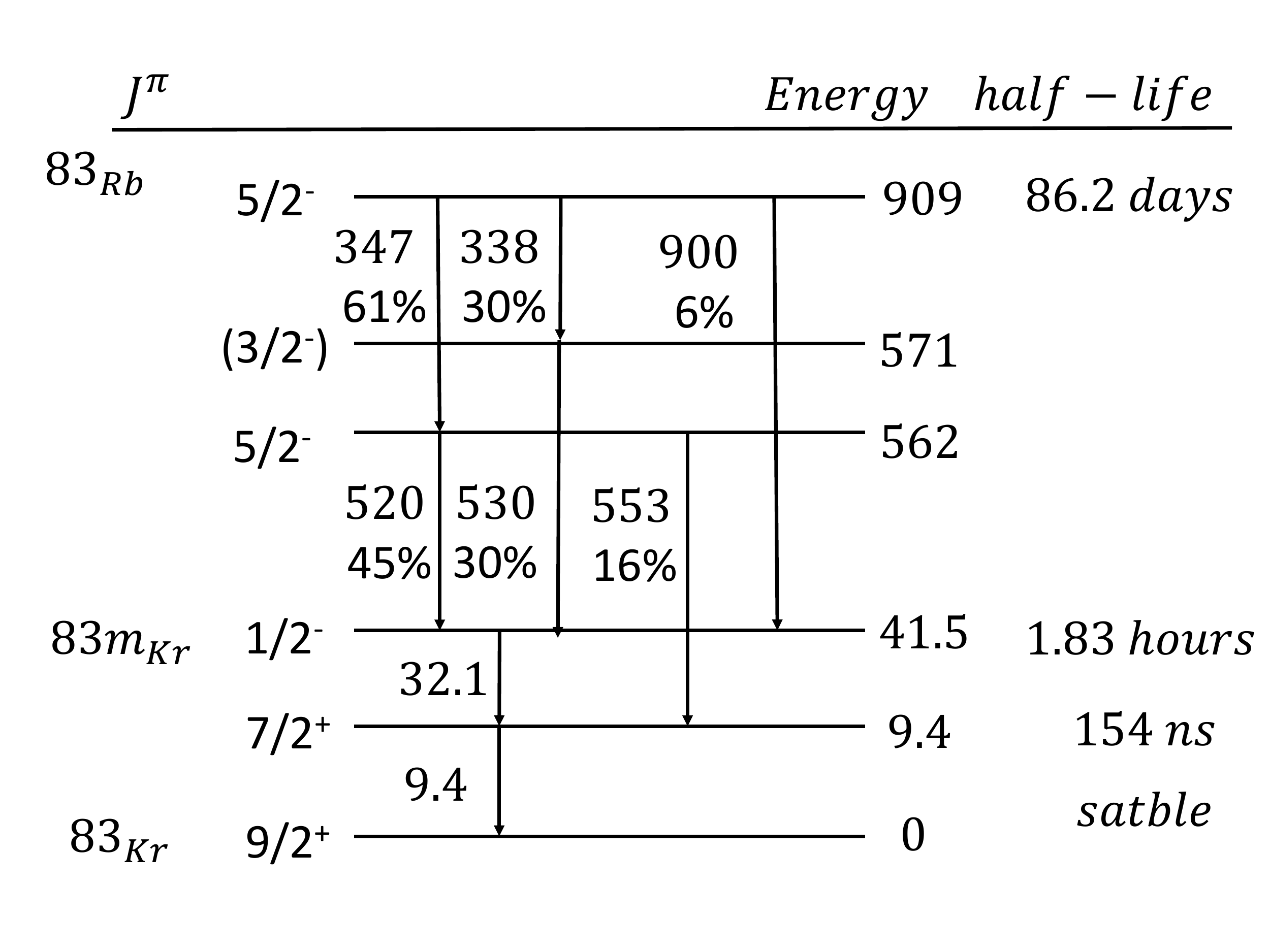}
\caption{\label{fig:kr_decay} Energy level diagram (in keV) for the $^{83}Rb$ decay. $^{83}Rb$ decays 75\% of the time to the long-lived isomeric $^{83m}Kr$ level that is 41.5keV above the ground state, which subsequently decays in two steps, a 32.1keV transition, typically a conversion electron and associated x-rays, followed by a similar 9.4keV transition~\cite{kr_xenon}.}
\end{centering}
\end{figure}

The isotopes $^{83m}Kr$ decays with a half-life of 1.83$\pm$0.02h via two electromagnetic transitions of energy 32.1 and 9.4 keV. $^{83m}Kr$ is produced in the decay of $^{83}Rb$ , which has a half-life of 86.2 days~\cite{Kr}, shown in Fig.~\ref{fig:kr_decay}. These decays in general result in the emission of a conversion electron accompanied by either an x ray or another electron. Both electrons and x rays produce electronic recoils in liquid noble gases, meaning that the full energy of the electromagnetic transitions participate in the production of scintillation light in the liquid via the same mechanism. As a noble gas, $^{83m}Kr$ gas can be introduced into detector through recirculation system without the need for a source insertion system and compromising the purity of the liquid. $^{83m}Kr$ has been used successfully in some large liquid noble gas detectors~\cite{DsKr,xenonKr}. $^{83m}Kr$ is also an important diagnostic tool for studying the $\beta$ spectrum of tritium to determine the neutrino mass in the KATRIN experiment~\cite{KATRIN}.

This paper describes successful tests of $^{83m}Kr$ as a calibration source in the liquid argon detector. $^{83m}Kr$ atoms adequately reach the center of the liquid volume to provide a good energy calibration. Besides, $^{22}Na$ is also used to calibrate the detector, as a comparison.

\section{Experiment Apparatus}
\begin{figure}[!htb]
\begin{centering}
\includegraphics[width=.45\textwidth]{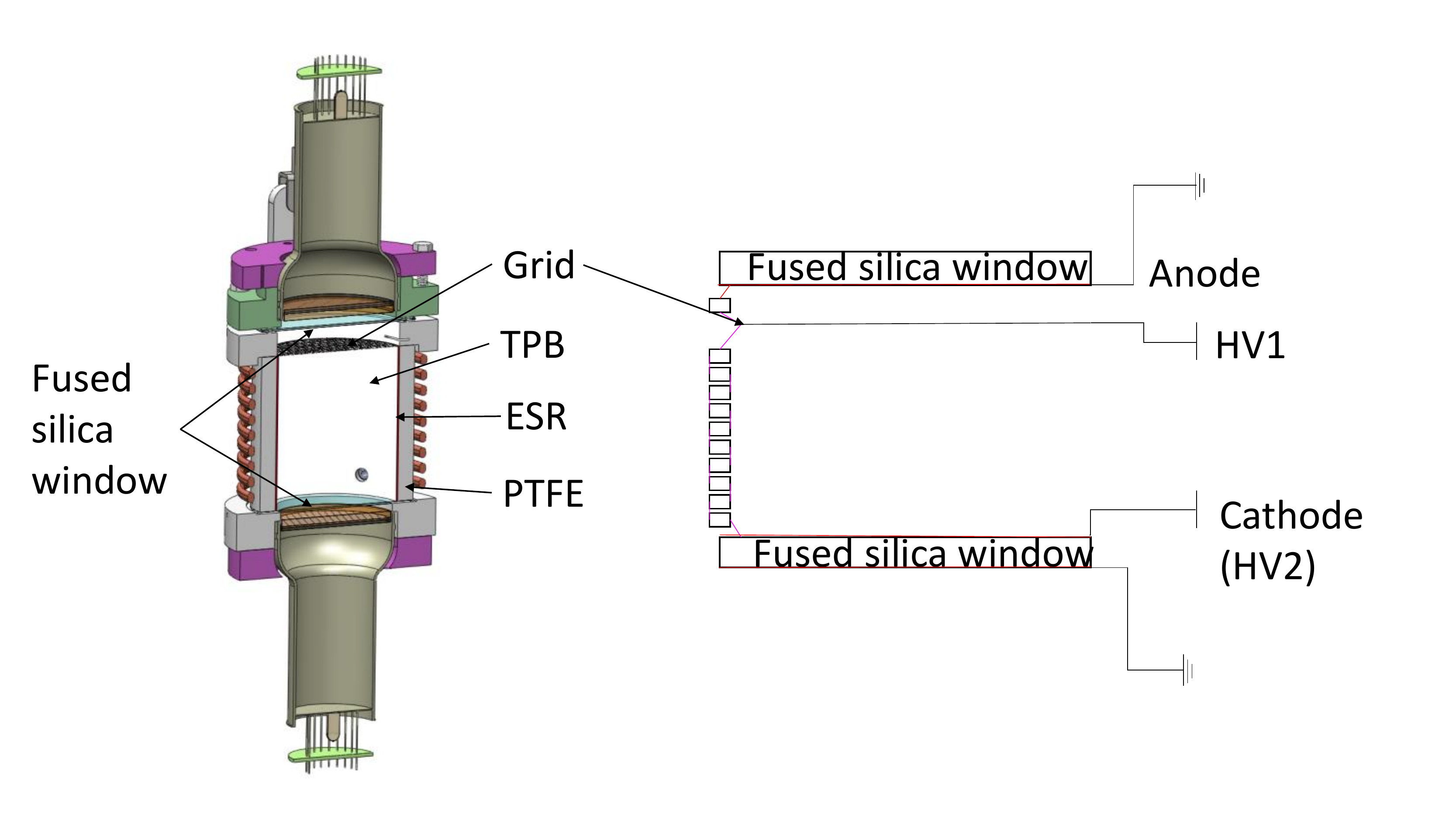}
\caption{\label{fig:detector} Left panel:Three-dimensional diagram of the detector. Right panel: the configuration of electric field. }
\end{centering}
\end{figure}

The liquid argon detector, shown in Fig.~\ref{fig:detector}, is a two-phase time projection chamber (TPC), which is constructed at institution of high energy physics, Chinese Academy of Sciences, in Beijing. The active liquid argon is about 8cm in diameter, 10.5cm in height, containing 0.74kg of liquid argon surrounded by 1cm thickness polytetrafluoroethylene (PTFE) sleeve. A layer of enhanced specular reflector film (ESR) is placed in the inner surface of the PTFE sleeve to increase reflectivity~\cite{ESR}. The silica glass windows 2mm thick, which are coated an transparent Indium Tin Oxide (ITO) film 15nm thick, placed at top and bottom, besides voltage divider made of a set of 4-mm thick copper rings around the PTFE sleeve are used to generate a uniform electric field throughout the active argon volume. The copper rings are connected by a kapton flexible printed circuit board, which is embedded with series resistor. An etched stainless steel grid is placed 5 mm below the liquid-gas surface to generate a strong extraction electric field. For TPC operation, the anode is grounded while the cathode and grid are independently-controllable voltages to set the drift and extraction fields. To shield the negatively-biased PMT photocathodes from the voltages applied to the cathode, the outer surface of bottom window is coated with a second ITO layer, which is grounded.

In order to detect the 128nm argon scintillation light, all inner surfaces include the ESR and windows are coated with wavelength shifter 1,1,4,4-tetraphenyl-1,3-butadiene (TPB) , which shifts the wavelength of the ultraviolet light to approximately 420nm~\cite{TPB}. The evaporative coating is performed in a 1.2m diameter and height high-vacuum chamber. The typical vacuum level reached prior to the evaporation is in the range of  4$\times10^{-6}Pa$ to 6$\times10^{-6}Pa$, TPB is heated to 230 degrees Celsius in a silica crucible. Two 3inch Hamamatsu PMTs(R11065), which are all immersed in the liquid argon, are used to collect the shifted scintillation light.

\begin{figure}[!htb]
\begin{centering}
\includegraphics[width=.45\textwidth]{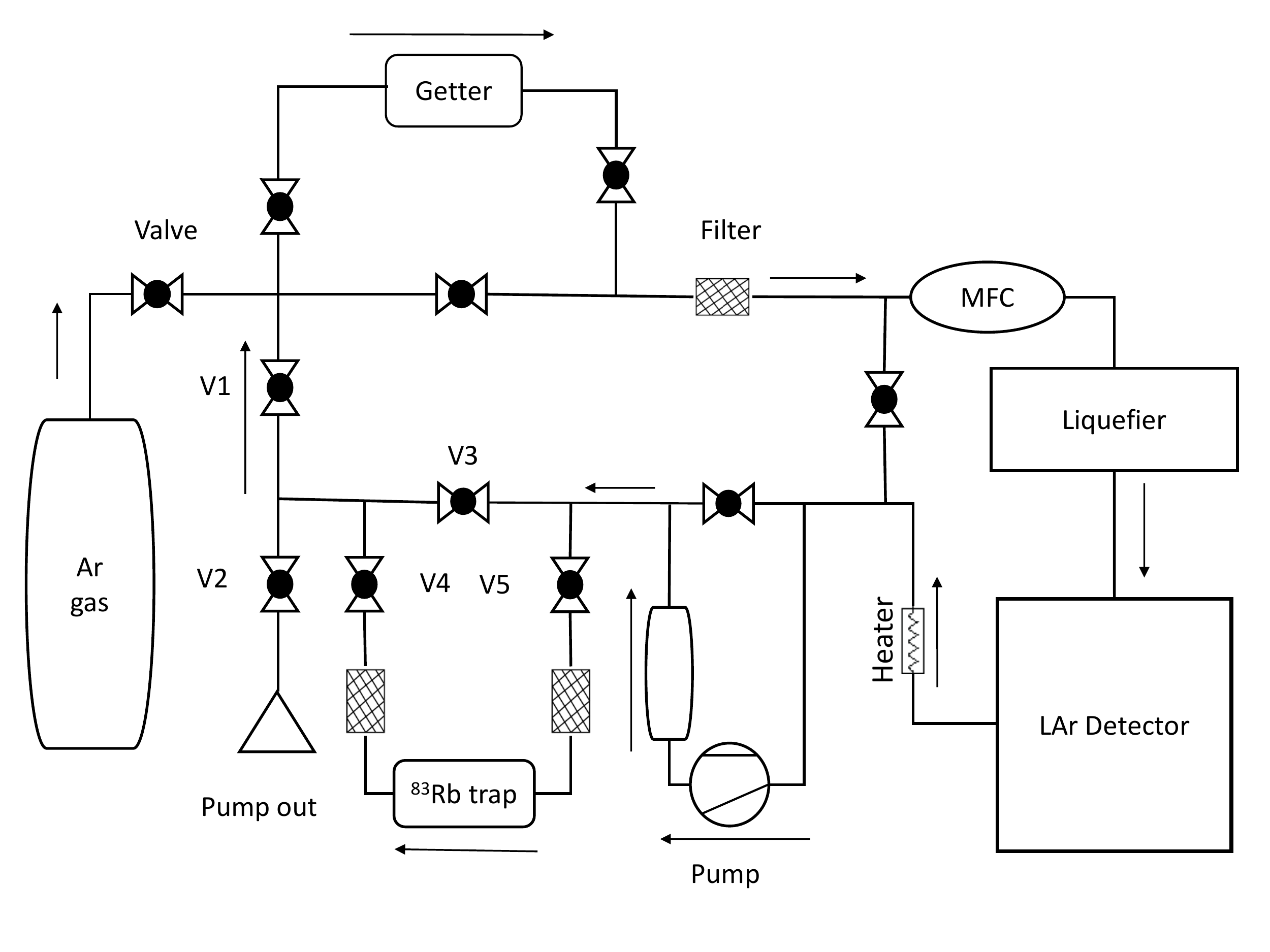}
\caption{\label{fig:gas-handling_system} The gas-handling system for the LAr detector. In $^{83m}Kr$ runs, valves V4 and V5 are open while valve V3 is closed. }
\end{centering}
\end{figure}

The $^{83}Rb$ radioactive source was produced at the institute of Modern Physics, Chinese Academy of Sciences, in LanZhou. The $^{83}Rb$ is infused in a 0.45g of zeolite, which absorbs $^{83}Rb$ but releases $^{83m}Kr$ gas, located in a gas cell called $^{83}Rb$ trap in the gas-handling system, shown in Fig.~\ref{fig:gas-handling_system}. The $^{83}Rb$ trap is connected to the gas inlet line outside the vacuum Dewar. In $^{83m}Kr$ tests, valve V4 is opened, V5 and V3 are closed. $^{83m}Kr$ gas is entrained in argon flow and introduced to the detector along with the circulating argon with a diaphragm pump about 20 l/min, While the argon gas flow through the $^{83}Rb$ trap. The filters prevent the introduction of zeolite into the argon purification system. Circulation then bypasses the $^{83}Rb$ trap, but still passes through the getter. The heater/circulation pump was activated to ensure the mixing of the liquid in the active region during $^{83m}Kr$ tests. In order to keep the LAr purity ,the circulating purification system has never been interrupted during the whole experimental process. The detail information of the cooling, purification and recirculation have been described in~\cite{LPX}.

%
%

%

\section{Data acquisition and single-photoelectron calibration}
\subsection{Data acquisition}
\begin{figure}[!htb]
\begin{centering}
\includegraphics[width=.45\textwidth]{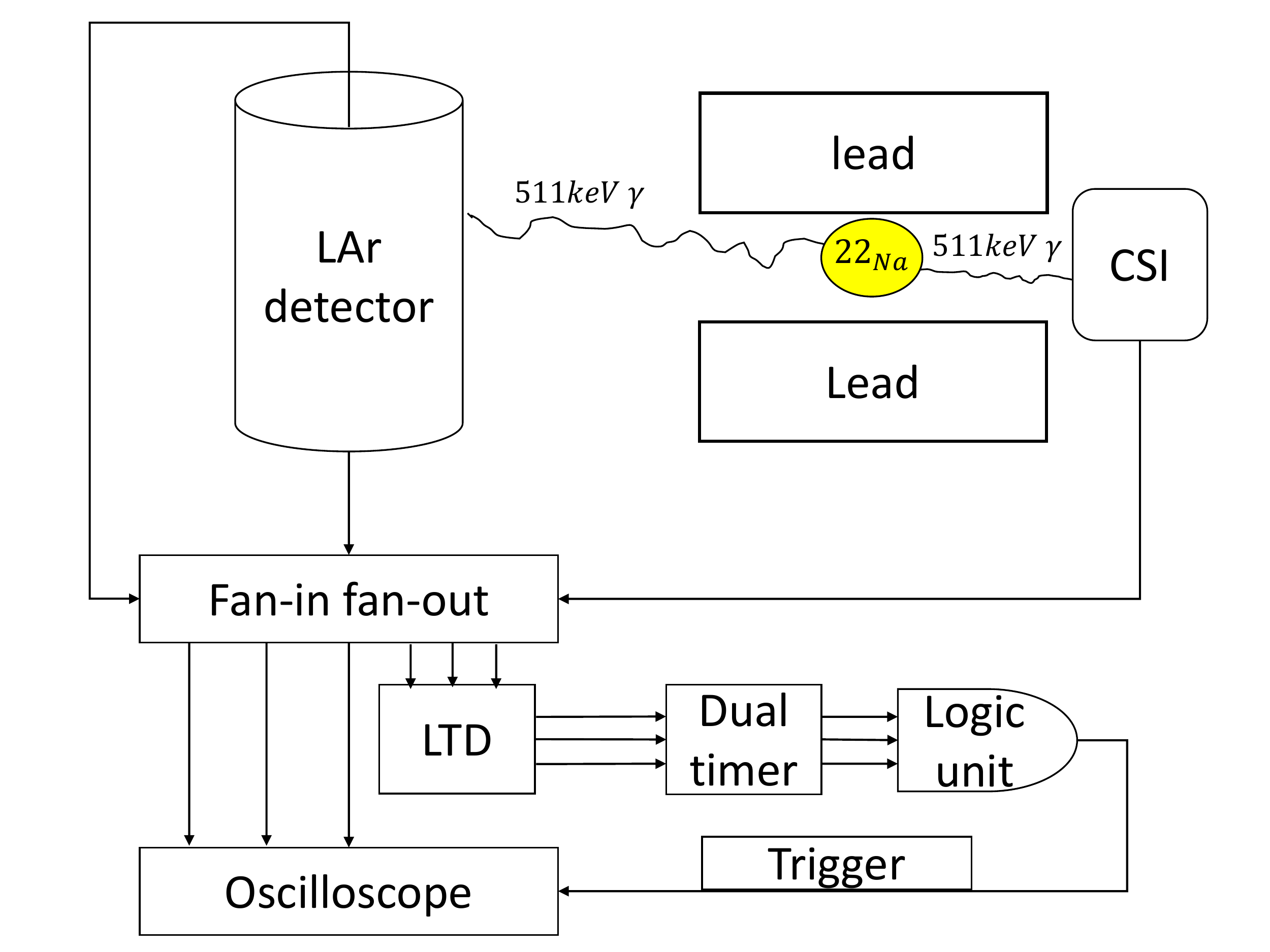}
\caption{\label{fig:DAQ} Logical diagram for the signal waveform recording in $^{22}Na$ run. The back-to-back $\gamma$-rays is tagged to measure the light yield.}
\end{centering}
\end{figure}

Signals from the PMTs are expands into two channels by a linear Fan in-Fan out. One is sent to a discriminator to form the  trigger and the other is sent to a LeCroy digitizing oscilloscope (1 sample per 2 ns, HDO6054) to record the signals waveform for off-line analysis. There are two trigger configurations. For $^{22}Na$ run, the trigger requires a coincidence of the top and bottom PMT signals in LAr, and the PMT signal of the CsI detector (A square CsI crystal with sides 8cm long coupled to a PMT) within 200ns. The threshold is set at 4mV for the PMTs in liquid argon, 7mV for the PMT of the CsI detector, respectively. The outputs of the discriminators are passed to a dual timer to produce a 10us signal, which is used to create a trigger signal in logic unit. The oscilloscope is set at 1us/division, 100mV/division and 10us time window (1 us before the trigger, 9 us after the trigger gate). Fig.~\ref{fig:DAQ} shows the logical diagram for the waveform signal record in $^{22}Na$ run. For $^{83}Kr^m$ run, only the PMTs signal in liquid argon are required for the trigger and the HDO6054 is set at 20mV/division.

\subsection{Single-photoelectron calibration}
\begin{figure}[!htb]
\begin{centering}
\includegraphics[width=.45\textwidth]{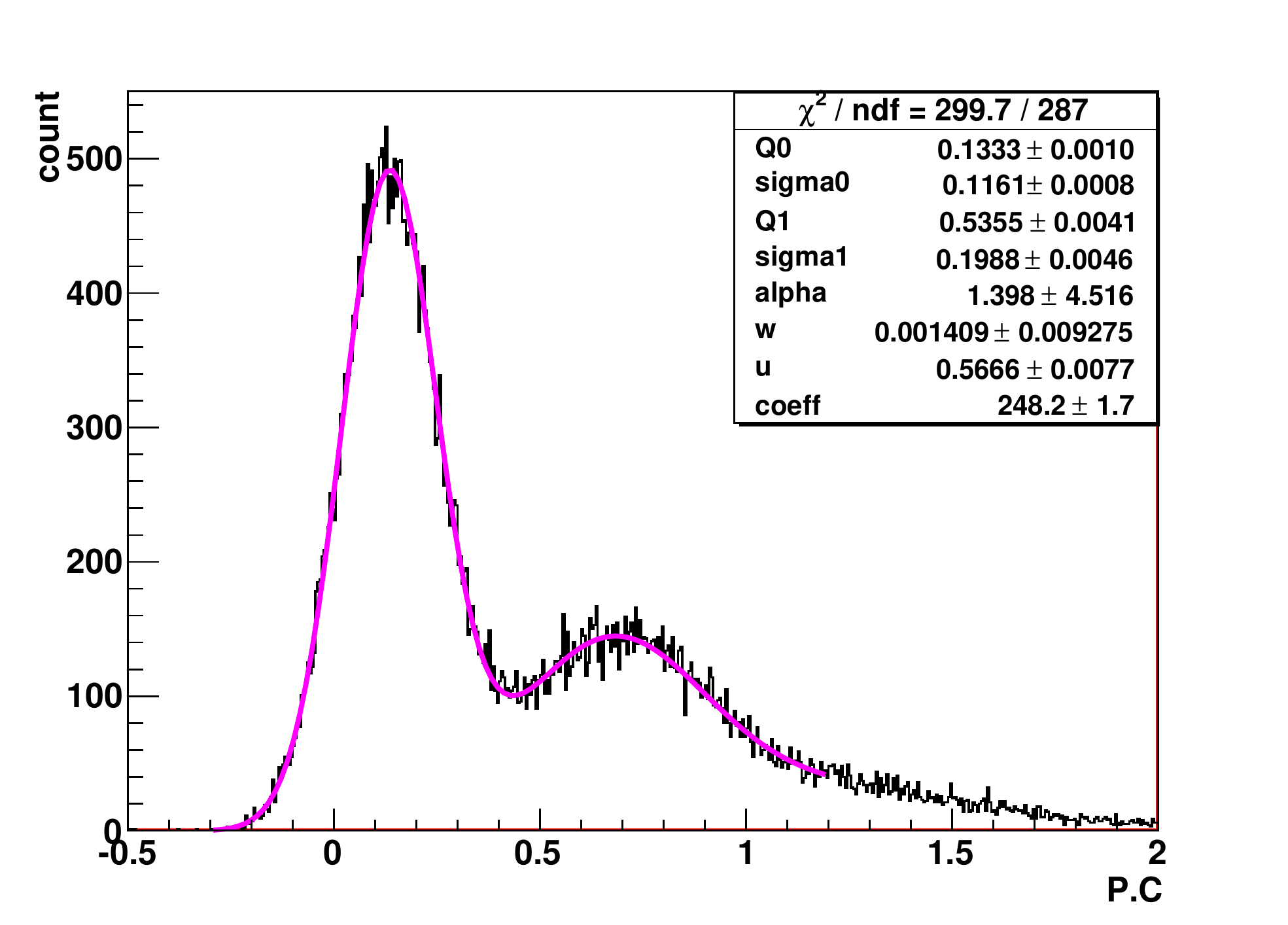}
\caption{\label{fig:spe_1750V} Example of the single photoelectron spectrum of a single PMT in the LAr.}
\end{centering}
\end{figure}

Single photoelectron spectra for the two PMTs are acquired by a LED installed in the detector, which is driven by a BNC8010 pulse generator. A sample spectrum for a LED calibration is shown in Fig.~\ref{fig:spe_1750V}. The spectrum is fitting using a PMT response function described in~\cite{spe}. The mean charges for the two PMTs are 0.53 pC and 0.44 pC. The single photoelectron distribution is monitored throughout the data-taking period with a variation throughout the running period of less than 2\%.

\section{Data analysis and results}
\label {Light yield}
\subsection{Light yield and $^{83m}Kr$ half-life }
\begin{figure}[!htb]
\begin{centering}
\includegraphics[width=.45\textwidth]{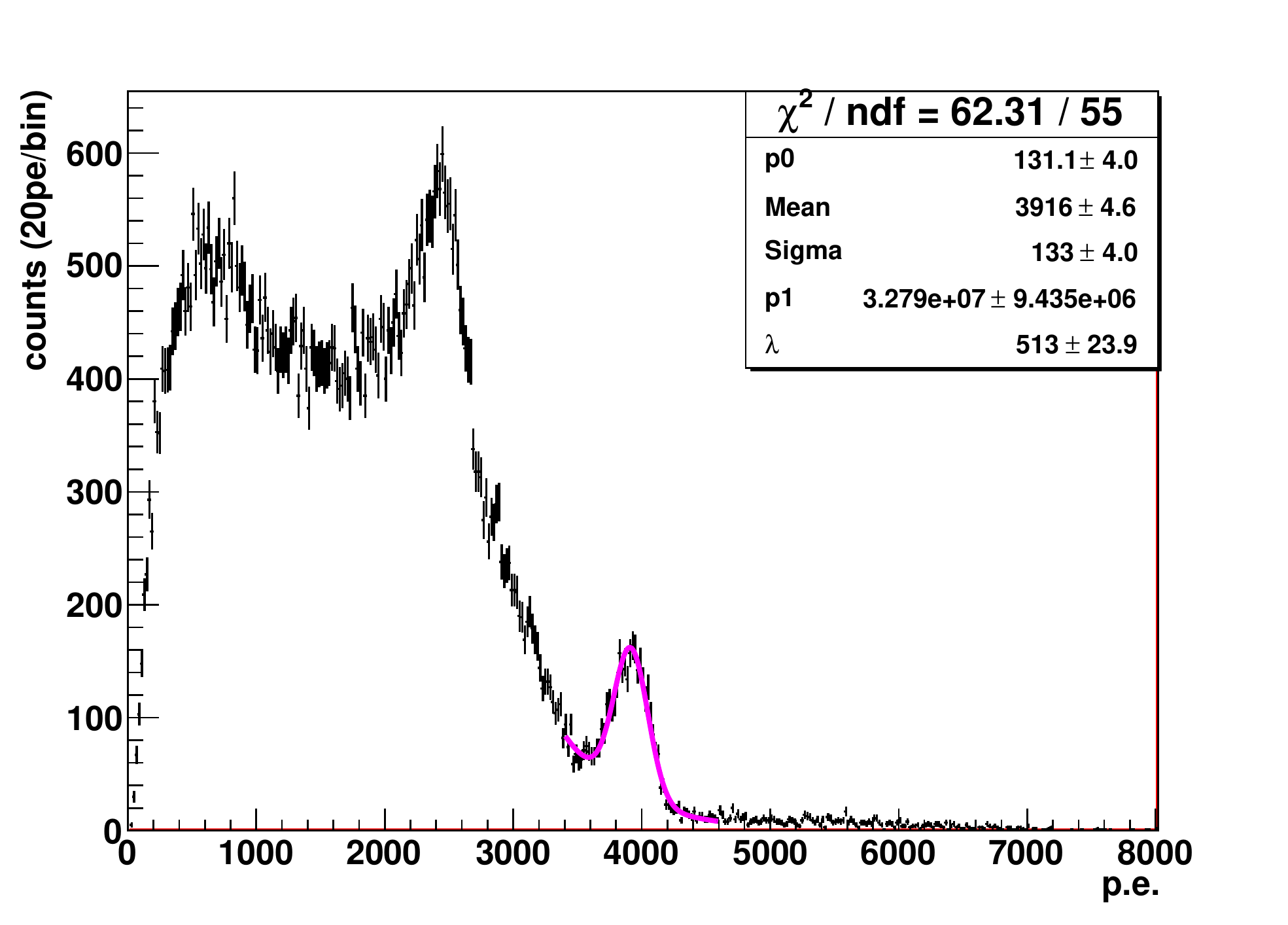}
\caption{\label{fig:511keV} Scintillation spectrum of $^{22}Na$ collimated at the central position, the data is not background subtracted because the background rate is negligible.}
\end{centering}
\end{figure}

The Liquid argon detector is calibrated firstly by a 40uCi encapsulated Na source placed outside the cryostat. The source emits a positron and a 1275 keV $\gamma$-ray. The positron annihilates in the encapsulation and produces two back to back 511 keV $\gamma$-rays. As is shown in Fig.~\ref{fig:DAQ} the $\gamma$-rays aims at the center of the active LAr target and is collimated by a 13-cm-thick lead collimator with a hole in 10mm diameter. The most prominent and reliable feature in the spectrum, namely the full energy peak, is used to calculate the light yield, shown in Fig.~\ref{fig:511keV}.

\begin{figure}[!htb]
\begin{centering}
\includegraphics[width=.45\textwidth]{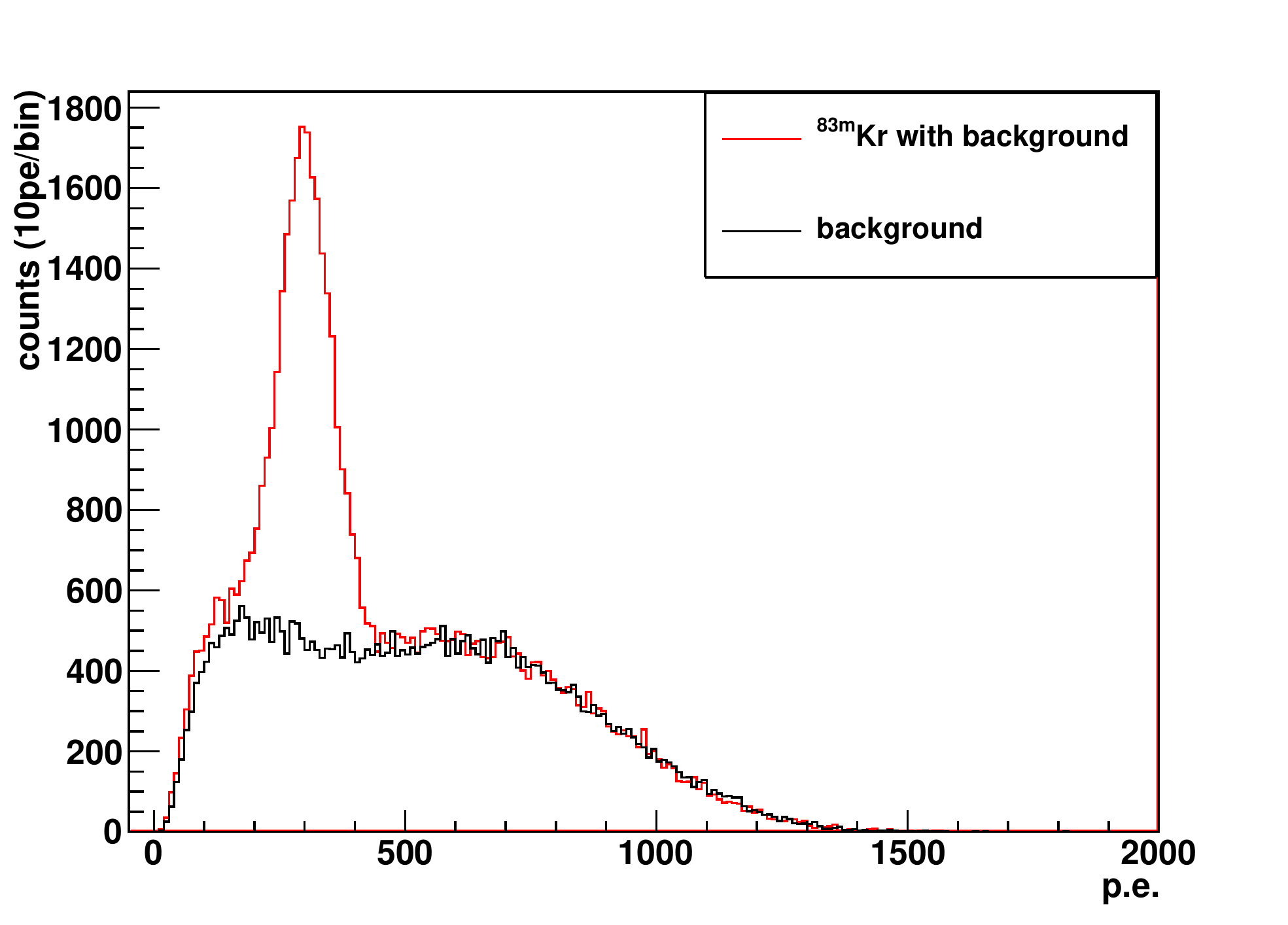}
\includegraphics[width=.45\textwidth]{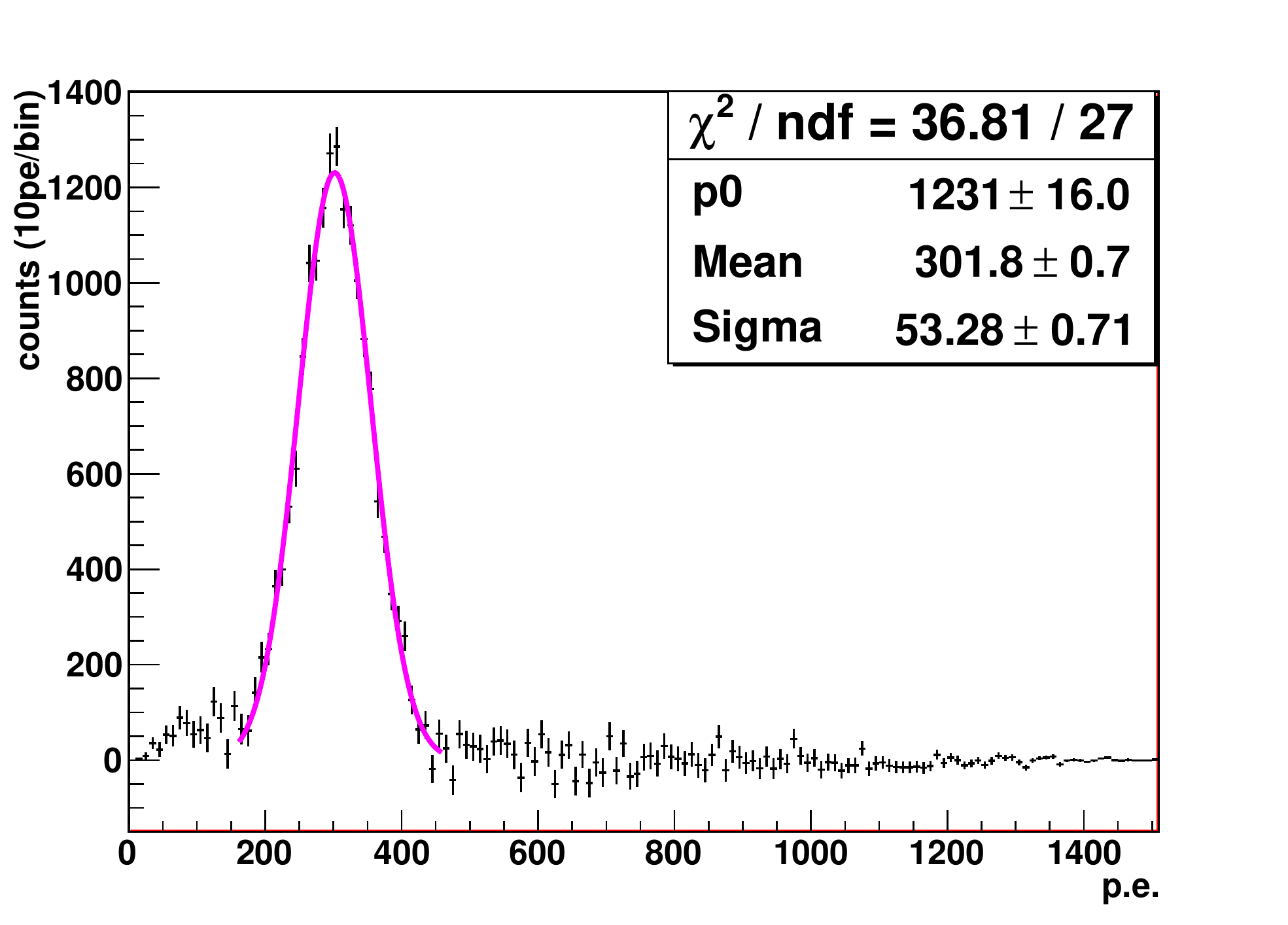}
\caption{\label{fig:kr_0V_2205}(top) The black and red curve indicate changes before and after $^{83}Kr^m$ is injected. (bottom)  Scintillation spectrum of $^{83}Kr^m$ of the full absorption peak with subtraction of a background.}
\end{centering}
\end{figure}

Before $^{83m}Kr$ run, several background runs have been taken for a background subtraction. The decay of $^{83m}Kr$ to stable $^{83}Kr$ is characterized by two closely-spaced sequential decays producing either IC electrons ,x-rays or $\gamma$s. In this case the ionization is produced by primary electrons but their short time separation ($\sim$ 154ns) cause the interactions to appear as a single one due to the long argon triplet scintillation time ($\sim$1.6us), shown in Fig.~\ref{fig:kr_0V_2205}. The Total energy released in the LAr is 41.5keV~\cite{Krpeak}.

\begin{figure}[!htb]
\begin{centering}
\includegraphics[width=.45\textwidth]{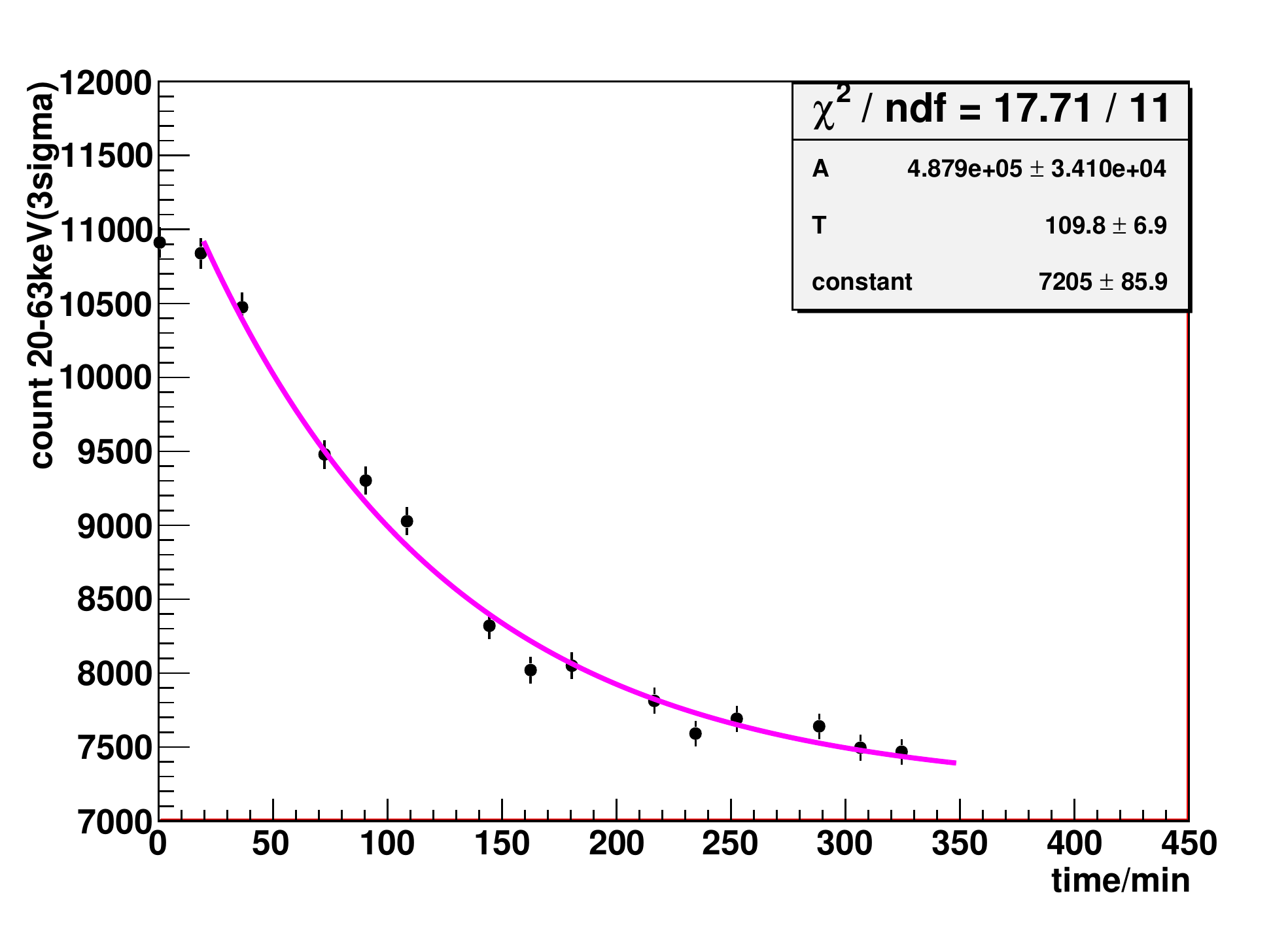}
\caption{\label{fig:kr_half-time} (color online) Rate of  $^{83m}Kr$ events between 20-63keV in LAr as a function of time from ending a fill. The rate decays with a fitted half-life of 109.8$\pm$6.9 minutes (1.83$\pm$0.11 h), consistent with the reported value of 1.83$\pm$0.02 h~\cite{Kr}.}
\end{centering}
\end{figure}

The signal baseline is determined before the trigger in acquisition window (10us gate wide), and then subtracted from the waveform. Two cuts are applied to remove from the spectrum. The amplitude cut is used to remove the signal beyond the scope of the oscilloscope and the time cut, namely the trigger time difference for two PMTs are not within 200ns, is used to remove the coincident events.

The light yield measurements have been performed  with the 511keV peak of $^{22}Na$ and 41.5keV peak of $^{83m}Kr$. A model function consisting of a Gaussian and a falling exponential is used to fit the full absorption peak at 511keV, the average light yield is approximately 7.66$\pm$0.01 photoelectrons/keV with an energy resolution of 3.6\%, shown in Fig.~\ref{fig:511keV}. For the 41.5keV peak from  $^{83m}Kr$, the energy spectrum is fitted with a gaussian function using a least- squares fitting method, shown in Fig.~\ref{fig:kr_0V_2205}, which returns a light yield of 7.28$\pm$ 0.02 photoelectrons/keV and 17.6\% resolution. The fitting results show the light of 511keV is about 5\% higher than that of 41.5keV, which is because the greater stopping power for lower energy $\gamma$-rays or electrons~\cite{stop_energy}.


After stop filling $^{83m}Kr$ atom, trigger rate decreases over time, reaching the background level in about 6 hours. Segmentation of 5 data sets was used to analyse the change of the count rate, which are limited to a energy range of 20keV to 63keV. Fig.~\ref{fig:kr_half-time} shows the $^{83m}Kr$ rate as a function time with a fitted half-life of 1.83$\pm$0.11 h, consistent with the reported value of 1.83$\pm$0.02 h~\cite{Kr}.


\subsection{light yield varies with the drift electric field from 0 to 200V/cm }
\label {Drift E}

\begin{figure}[!htb]
\begin{centering}
\includegraphics[width=.45\textwidth]{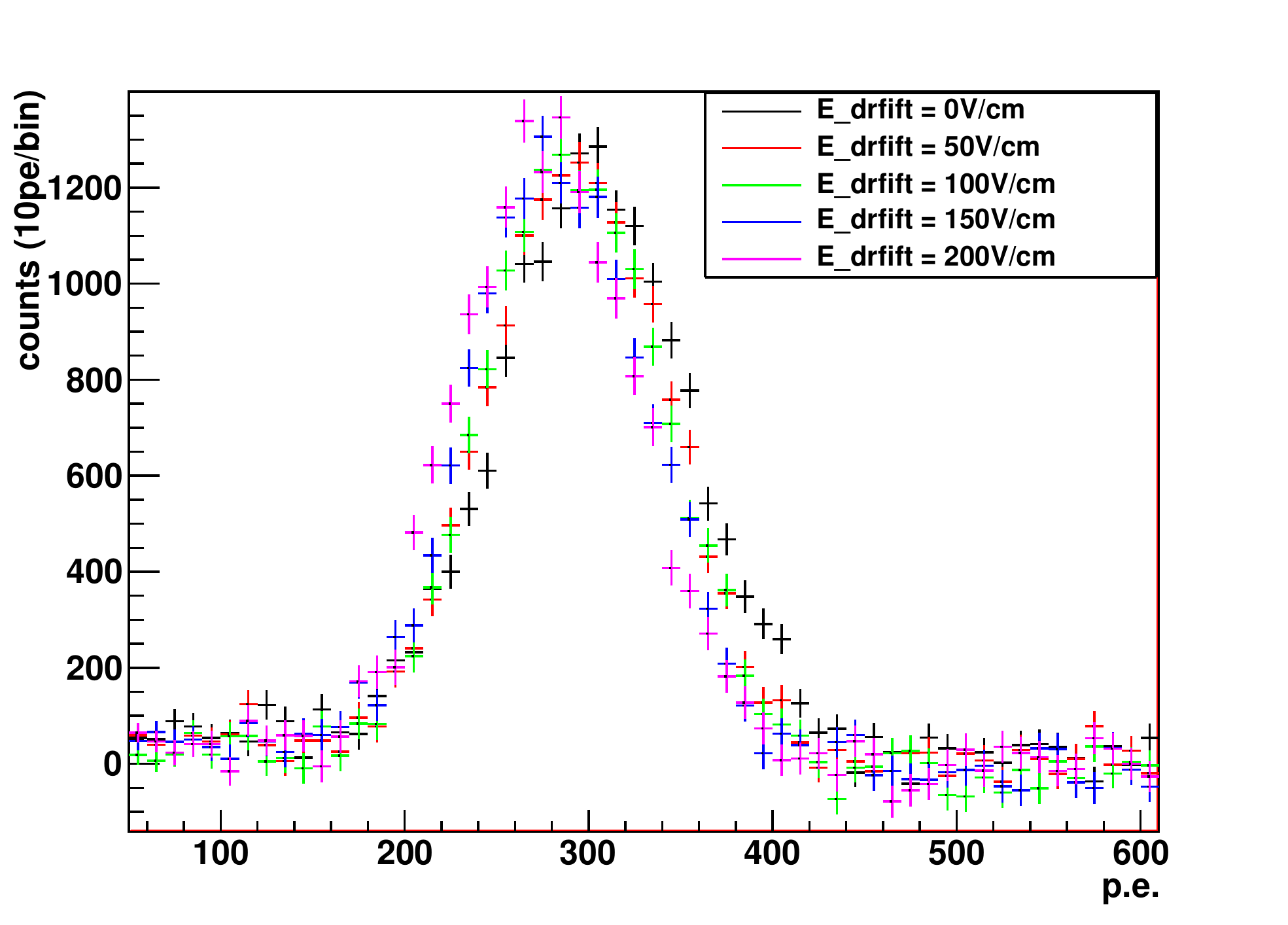}
\caption{\label{fig:Kr_0to200V}  The full-absorption peak of $^{83m}Kr$ at all drift fields, the fit results are listed in Table~\ref{table:Kr}}
\end{centering}
\end{figure}

\begin{figure}[!htb]
\begin{centering}
\includegraphics[width=.45\textwidth]{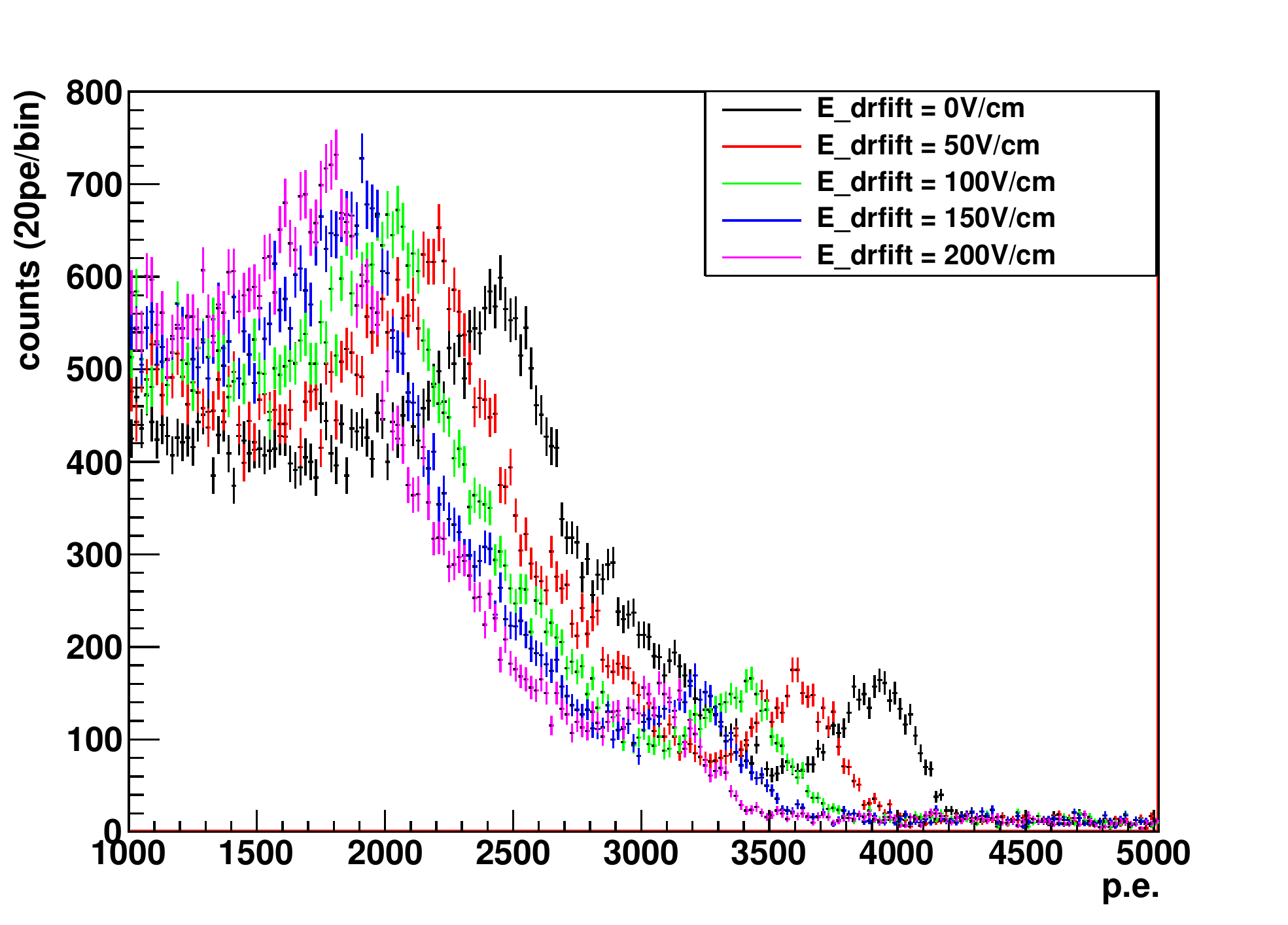}
\caption{\label{fig:Na_0to200V}  The 511keV full-absorption peak of $^{22}Na$ at all drift fields,the fit results are listed in Table~\ref{table:Na}.}
\end{centering}
\end{figure}

\begin{table}[tb]
  \caption{Fitted $^{83m}Kr$ full-absorption peak mean, width,and light yield varies with the drift electric field from 0 to 200V/cm. The error on $\mu_{p}$ and $LY$ is the statistical error from the fit.}
  \label{table:Kr}
  ~\\
  \begin{tabular}{cccc}
    \hline
  \textbf{$E[V/cm]$} & \textbf{$\mu_{p}[pe]$} & \textbf{$\sigma_p[pe]$} & \textbf{ $LY[pe/keV_{ee}]$}\\
    \hline
  0 & 301.9$\pm$0.7 &   52.88   &   7.28$\pm$0.02 \\
  50 & 294.0$\pm$0.7 &   49.72   &  7.08$\pm$0.02 \\
  100 & 291.3$\pm$0.7 &  47.81    &  7.02$\pm$0.02 \\
  150 & 283.1$\pm$0.7 &   49.15   &  6.82$\pm$0.02 \\
  200 & 276.9$\pm$0.7 &   48.44  &  6.67$\pm$0.02 \\
    \hline
  \end{tabular}
\end{table}

\begin{table}[tb]
  \caption{Fitted $^{22}Na$ full-absorption peak mean, width,and light yield varies with the drift electric field from 0 to 200V/cm. The error on $\mu_{p}$ and $LY$ is the statistical error from the fit.}
  \label{table:Na}
  ~\\
  \begin{tabular}{cccc}
    \hline
  \textbf{$E[V/cm]$} & \textbf{$\mu_{p}[pe]$} & \textbf{$\sigma_p[pe]$} & \textbf{ $LY[pe/keV_{ee}]$}\\
    \hline
  0 & 3916$\pm$4.6 &   133.0   &   7.66$\pm$0.01 \\
  50 & 3607$\pm$5.7 &   142.7   &  7.06$\pm$0.01 \\
  100 & 3390$\pm$6.1 &   147.5   &  6.63$\pm$0.01 \\
  150 & 3220$\pm$5.9 &   149.5   &  6.30$\pm$0.02 \\
  200 & 3078$\pm$6.9 &   1475.3  &  6.02$\pm$0.02 \\
    \hline
  \end{tabular}
\end{table}

Fig.~\ref{fig:Kr_0to200V} and Fig.~\ref{fig:Na_0to200V} show the light yield varies with the drift electric field intensity. When a particle interacts in noble gases it produces excited (excitons) and ionized (ions) atoms with soft elastic recoils which eventually thermalise as heat. Excitons then decay via the formation of an excited dimer , producing photons, the ions need to undergo recombination process to form an excited dimer. In the presence of an electric field, some of the ionized electrons can escape the electron-ion cloud, reducing recombination effect, which results in a decrease in light yield ~\cite{Kubota}. The decrease of light yield is obvious and increases as the electric field intensifies, shown in Table~\ref{table:Kr} and Table~\ref{table:Na}. The data includes at light yield results with drifting field of 50V/cm, 100V/cm, 150V/cm,and 200V/cm along with null field.

\section{Conclusions}

The ability, that $^{83m}Kr$ atoms are successfully introduced and detected in the liquid argon, has been demonstrated. $^{83m}Kr$ decays in a few hours to stable $^{83}Kr$, which is an excellent calibration source to characterize the scintillation signal yield of liquid argon detector at low energy and not create permanent contamination. Besides, $^{83m}Kr$ can be used to calibrate the energy scale in the fiducial region and monitor detector stability.


The variation of the light yield with the drift electric field from 0 to 200V/cm have been measured. The $^{83m}Kr$ and $^{22}Na$ results have been shown in Table~\ref{table:Kr} and Table~\ref{table:Na}. The PARIS~\cite{PARIS} model provides a good description of electric recoils (ERs) in liquid argon. The measured electric field induced quenching of light yield for ERs is 8.4\% for $^{83m}Kr$ and 21.4\% for $^{22}Na$ at 200V/cm. The difference of quenching factor is because of the recombination probability decreases with the increase of the particle energy in the range of 40keV to 511keV, which is consistent with Ref.~\cite{ARIS}.

\section{Acknowledgments}

We acknowledge financial support from supported by Ministry of Science and Technology of the People's Republic of China (2016YFA0400304).
 We would like to thank Institute of Modern Physics of the Chinese Academy of Sciences and Shanghai jiao tong university for the support of the production of the $^{83}Rb$ source. We also thank Y. Wang, a postdoctoral fellow at UCLA helped me in the early days of the detector design.

\end{document}